\documentclass[openacc]{rsproca_new}

\usepackage{graphicx}
\usepackage{dcolumn}
\usepackage{bm}
\usepackage{color}
\usepackage{mathtools}
\usepackage{amsmath}
\usepackage{hyperref}

\def\cFrac#1#2{%
\begin{array}{@{}c@{}}\multicolumn{1}{c|}{#1}\\%
\hline\multicolumn{1}{|c}{#2}\end{array}}

\begin{document}

\title{Regulation of stem cell dynamics through volume exclusion}

\author{
Rodrigo Garc{\'i}a-Tejera$^{1,2}$, Linus Schumacher$^{1,2}$ and Ramon Grima$^{2}$}

\address{$^{1}$Centre for Regenerative Medicine, University of Edinburgh, 5 Little France Dr, Edinburgh EH16 4UU, U.K.\\
$^{2}$School of Biological Sciences,
Kings Buildings, Mayfield Road, \\ University of Edinburgh, EH9 3JF, U.K.}

\subject{mathematical physics, statistical physics \\

\textbf{Cite this article:} García-Tejera R, Schumacher
L, Grima R. 2022 Regulation of stem cell
dynamics through volume exclusion. Proc. R.
Soc. A 478: 20220376. \\
\href{https://doi.org/10.1098/rspa.2022.0376}{https://doi.org/10.1098/rspa.2022.0376}\\
\vspace{0.5 cm}
\textbf{\textit{Received:}} 1 June 2022 \\
\textbf{\textit{Accepted:}} 27 September 2022}

\keywords{stem cells, volume exclusion, master equation,
renormalized system-size expansion, transient
bimodality, competition for space}

\corres{Rodrigo Garc{\'i}a-Tejera, \\
\email{rodrigo.garcia@ed.ac.uk} \\
https://orcid.org/0000-0003-3427-3587}

\begin{abstract}
Maintenance and regeneration of adult tissues rely on the self-renewal of stem cells. Regeneration without over-proliferation requires precise regulation of the stem cell proliferation and differentiation rates. The nature of such regulatory mechanisms in different tissues, and how to incorporate them in models of stem cell population dynamics, is incompletely understood. The critical birth-death (CBD) process is widely used to model stem cell populations, capturing key phenomena, such as scaling laws in clone size distributions. However, the CBD process neglects regulatory mechanisms. Here, we propose the birth-death process with volume exclusion (vBD), a variation of the birth-death process that considers crowding effects, such as may arise due to limited space in a stem cell niche. While the deterministic rate equations predict a single non-trivial attracting steady state, the master equation predicts extinction and transient distributions of stem cell numbers 
with three possible behaviours: long-lived quasi-steady state, and short-lived bimodal or unimodal distributions. In all cases, we approximate solutions to the vBD master equation using a renormalized system-size expansion, quasi-steady state approximation and the WKB method. Our study suggests that the size distribution of a stem cell population bears signatures that are useful to detect negative feedback mediated via volume exclusion.
\end{abstract}



\maketitle 

\section{\label{sec:Introduction}Introduction}

Stem cells (SCs) are a population of cells capable of self-renewing and differentiating into all the cells in a particular lineage. In adult tissue homeostasis, SCs slowly self-renew and differentiate to compensate for the death of other cells while maintaining a constant average population size \cite{till1964stochastic,tothova2007foxo,morrison2014bone,north2007prostaglandin}. Upon injury, however, the SC proliferation and differentiation rates increase dramatically to repair the tissue, only settling back into homeostasis when regeneration is completed \cite{becker2019stem,lander2009cell,renardy2018control}. Such tight control over the SC proliferation and differentiation rates requires regulatory mechanisms providing feedback to the SCs.    

Against the backdrop of recent advances in experimental techniques in stem cell biology, a broad range of SC regulatory mechanisms have been reported, such as negative feedback exerted by the more differentiated cells \cite{renardy2018control,lander2009cell,rodriguez2013stem}, competition between SCs for fate determinants \cite{yoshida2018open,kitadate2019competition}, or mechanical feedback \cite{hannezo2016interplay,vining2017mechanical,zhang2021phase}. An additional plausible mechanism stems from the confinement of SCs to a particular microenvironment, the stem cell niche  \cite{tumbar2004defining,zhang2003identification,li2005stem,yoshida2018open}. This microenvironment plays a key role in maintaining cell stemness and promoting SC quiescence \cite{so2018molecular,sampath2018induction,arai2008quiescent}, self-renewal, or differentiation, according to tissue requirements; however, it also triggers a competition between SCs for niche access \cite{stine2013stem,corominas2020stem,kitadate2019competition}.         

Crowding effects are often associated with volume exclusion \cite{smith2017macromolecular,wilson2017reactions,dyson2015importance,flory1966effect,segall2006volume}. The non-negligible volume of particles restricts their movement, thus obstructing their access to available free space \cite{smith2017macromolecular}. As a consequence, the accessible phase space can be greatly reduced. If cells are dividing without reducing their size, crowding effects can have an impact on the SC proliferation and death (or differentiation) rates. For example, a proliferation event reduces the available space, which in turn reduces the proliferation rate, thus creating a negative feedback loop. Volume exclusion has been also suggested to play a role in the regulation of cancer stem cells and tumour growth \cite{hillen2013tumor}. However, it is not yet clear how to distinguish between crowding effects and other regulatory mechanisms from observations of the population evolution (e.g. from snapshots of the SC population at different times).

Stem cell division and differentiation has been previously modelled stochastically by the simple chemical reaction network $S \rightarrow 2S$, $S \rightarrow \emptyset$ \cite{till1964stochastic,klein2011universal}, where differentiation is equivalent to death if differentiated progeny do not self-renew. To prevent the population from diverging or vanishing, the birth and death rates must be equal, thus obtaining a critical birth-death process (CBD). For stem cell population dynamics, the CBD process has been frequently treated under well-mixed and dilute gas conditions, which facilitate its computational implementation, e.g., Gillespie algorithm, \cite{gillespie1976general,schnoerr2015comparison} and mathematical analysis through the master equation formalism \cite{harris1963theory,grima2015linear}. This approach has been successfully employed to illustrate key features of SC populations, such as population asymmetry (the maintenance of a constant average population via symmetric divisions that are balanced at the population level, instead of asymmetric divisions), neutral competition \cite{till1964stochastic,klein2011universal,simons2011strategies}, and scaling properties of clone size distributions \cite{simons2011strategies,klein2011universal,yamaguchi2017dynamical,robertson2021longitudinal}. However, the CBD ignores the finite-size nature of cells and thus disregards the role of available space in cell division. 

Here, we present a modification of the birth-death process that includes competition for niche access, the birth-death process with volume exclusion (vBD). We subdivide the space within a niche into $N$ voxels (small volumes of space); each voxel is either occupied by a stem cell or else is empty. Assuming well-mixed conditions, the effective chemical reaction network describing this process is
$S+E \rightarrow 2S$, $S \rightarrow E$, where $S$ and $E$ describe stem cells and empty voxels in the niche, respectively. The first reaction reflects the need for a stem cell to find an empty voxel to divide, while the second one represents the birth of an empty voxel after stem cell death or differentiation (assuming differentiated progeny leave the niche space). Naturally the system obeys the conservation law $n_S+n_E=N$, where $N$ is the niche carrying capacity, and $n_S$, $n_E$ are the number of stem cells and empty voxels, respectively. Note that the vBD resembles a stochastic SIS model, with the number of infectious given by the species $S$, and susceptible by $E$. However, while the SIS model is usually treated for $N\rightarrow \infty$, we are interested in the low $N$ behaviour. In terms of the vBD model parameters, the basic reproduction number for the equivalent SIS model is given by $R_0=Nk_1/k_2$, where $k_1$ and $k_2$ are the rates of proliferation and differentiation, respectively. The deterministic rate equations for the vBD process predict a logistic convergence to a non-trivial attracting steady state. From a microscopic perspective (i.e. the master equation's solutions), however, this prediction is not realised, and the vBD process relaxes to extinction, irrespective of the parameter values and initial conditions. 

Our analysis reveals the three different behaviours of the stochastic vBD process that are absent in its deterministic counterpart. When the birth rate is much larger than the death rate, the system quickly takes the form of a long-lived, quasi-steady state, and very slowly relaxes to extinction through a transient bimodal distribution. Conversely, for death rates much larger than the birth rates, the system quickly converges to extinction through a unimodal transient. 
Lastly, when the birth and death rates are comparable, the transient distribution is bimodal but the convergence to extinction is fast. For these three different parameter regimes, we approximate the solution of the vBD master equation using a quasi-stationary approximation, a renormalized system-size expansion (including finite size corrections to the linear-noise approximation), and the WKB method. In particular, the renormalized system-size expansion is a recent modification of the original van Kampen's SSE that has not been widely used yet, but proves useful for tackling master equations of non-linear birth-death processes. Finally, we derive an expression for the expected extinction time, based on Kolmogorov's backward equation and first-passage time theory. Our analytical solutions provide insights into the rich behaviours of the vBD model. 

\section{\label{sec:Model} $\textrm{v}$BD model}
\begin{figure*}
    \begin{center}
        \includegraphics[scale=0.49]{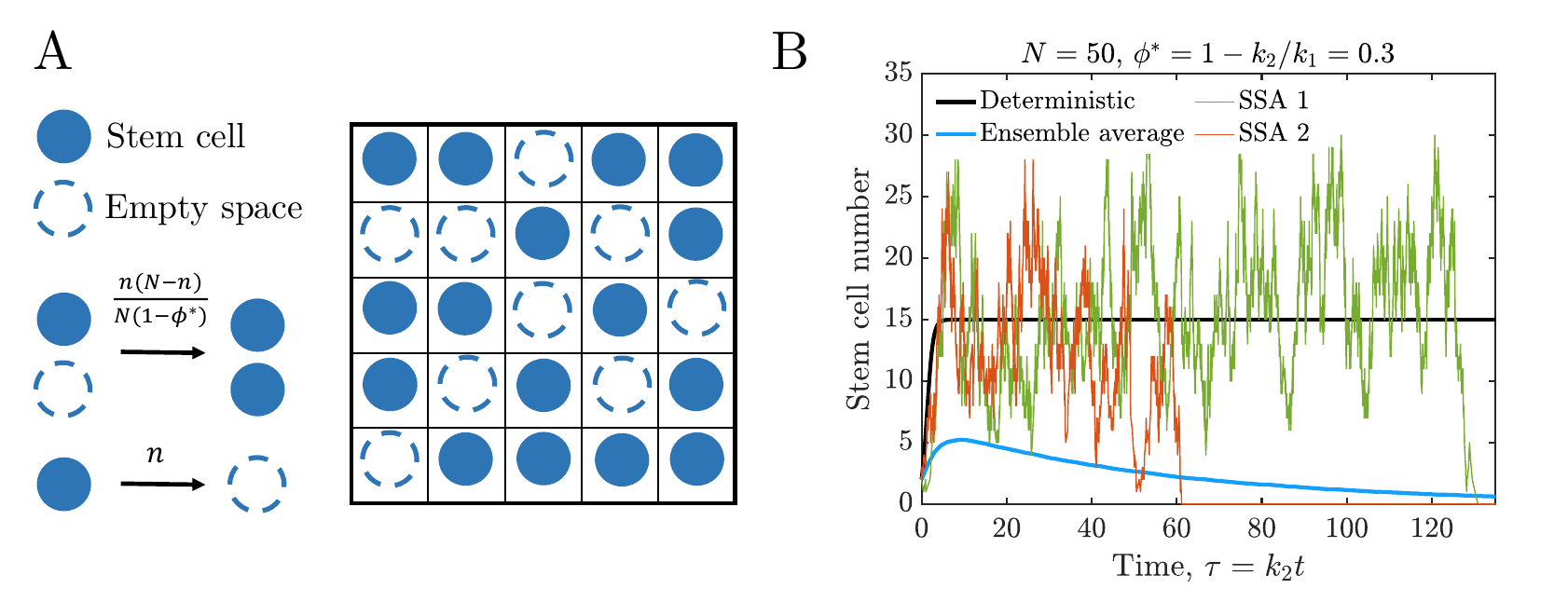}
    \end{center}
    \caption{The vBD model and its deterministic approximation. \textbf{A} Stem cells (solid circles) move randomly in the niche (grid) by switching position with empty voxels (dashed, empty circles). At any time a stem cell can ``react'' with a neighbouring empty voxel to create two stem cells (cell division), and a single stem cell can differentiate or die, leaving a new empty voxel in return. The propensities follow mass-action kinetics. \textbf{B} Under well-mixed conditions, the deterministic approximation predicts the evolution of the mean stem cell concentration satisfying the logistic equation \eqref{eq:meanFieldEquation}, thus portraying a logistic convergence to the average stem cell number $N\phi^*$ (black line). Stochastic trajectories (green and orange lines) obtained by the SSA, however, are driven to the extinction state by fluctuations. The ensemble average (blue line) of $2 \times 10^4$ stochastic trajectories reveals the eventual distinction--in contrast to the deterministic prediction of a non-zero steady state.  
    \label{fig:meanFieldApproach}} 
\end{figure*}

The birth-death process with volume exclusion is defined by the chemical reaction network
\begin{equation}
    S+E \xrightarrow{k_1} 2S; \qquad S \xrightarrow{k_2} E, 
    \label{eq:vBD_CRN}
\end{equation}
\noindent where $S$ and $E$ represent stem cells and empty voxels, respectively (See Fig.~\ref{fig:meanFieldApproach}{\bf A} for an illustration). For the deterministic system to have a non-trivial steady state, we require that $k_1>k_2$. Note that the two species are coupled by the conservation law $n_S+n_E=N$. Assuming mass-action kinetics and defining the dimensionless time $\tau=k_2 t$, the rate equation for the average stem cell concentration $\phi = n_S/N$ adopts the logistic form 
\begin{equation}
\frac{\partial \phi}{\partial \tau} = \frac{(\phi^* - \phi) \phi}{1-\phi^*},     
\label{eq:meanFieldEquation}
\end{equation}
\noindent where $\phi^*=1-k_2/k_1$ is the non-trivial steady state. The deterministic evolution of the stem cell concentration has the form $\phi(\tau)= (\phi^* \phi_0) / [(\phi^*-\phi_0) e^{- \phi^* \tau/(1-\phi^*)} + \phi_0]$, where $\phi(0)=\phi_0$ is the initial condition, and we can appreciate that $\phi \xrightarrow{\tau \rightarrow \infty} \phi^*$ when $\phi_0 \neq 0$ (Fig.~\ref{fig:meanFieldApproach}{\bf B}). In this deterministic system, the extinction state $\phi=0$ is never reached unless $\phi(0)=0$. 

The stochastic behaviour of the vBD differs from the deterministic predictions. Trajectories of the vBD generated using the stochastic simulation algorithm (SSA \cite{gillespie1976general}) fluctuate in the vicinity of the deterministic steady state for some finite period of time (Fig.~\ref{fig:meanFieldApproach}{\bf B}), but fluctuations eventually drive the stem cell number to extinction. The ensemble average of stochastic trajectories thus converges to zero (see blue line in Fig.~\ref{fig:meanFieldApproach}{\bf B}), disagreeing with the deterministic model's prediction.  

\begin{figure*}
    \begin{center}
        \includegraphics[scale=0.44]{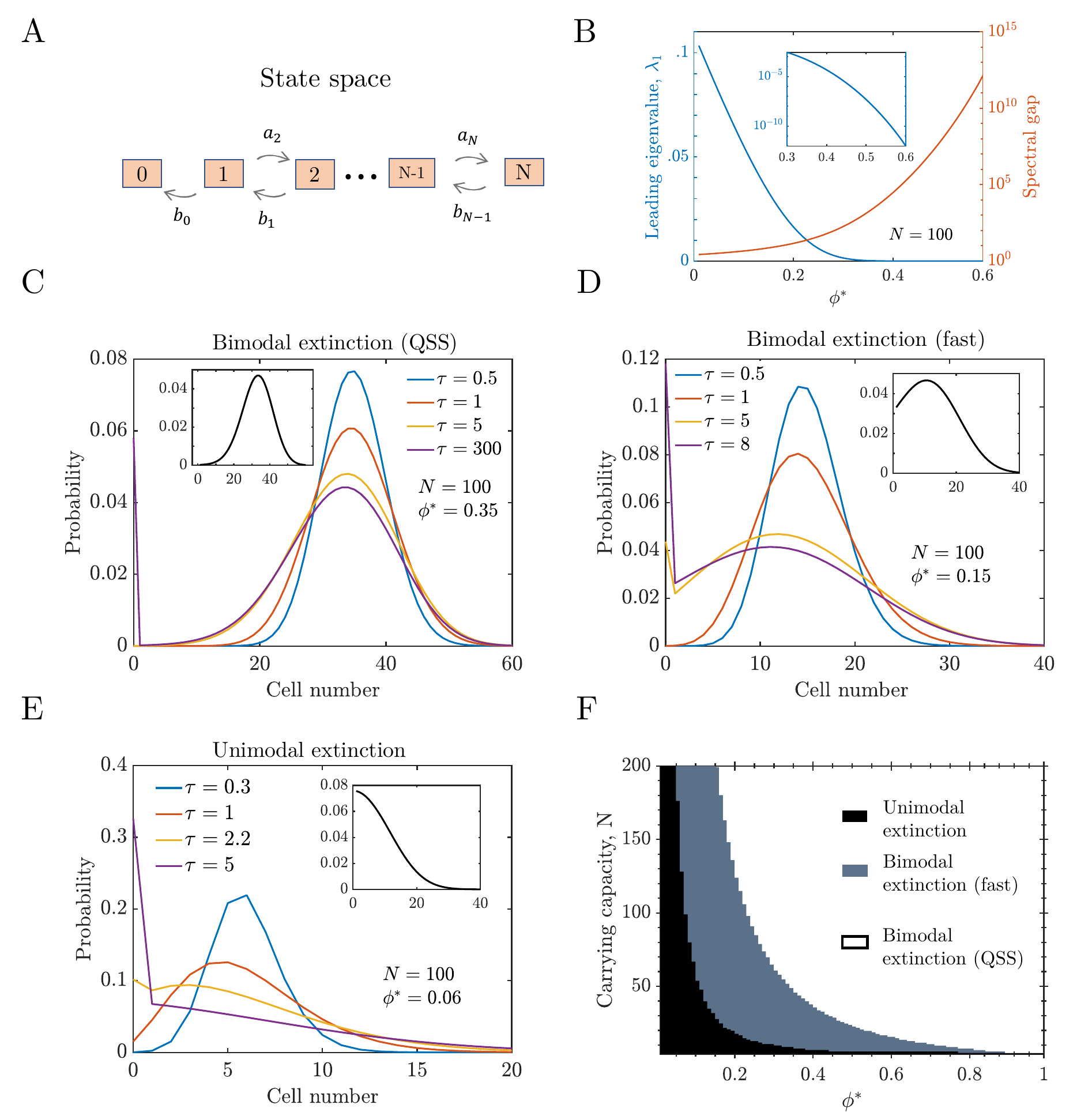}
    \end{center}
    \caption{Stochastic properties of the vBD model. \textbf{A} The vBD system jumps between states of $1,2,\dots,N$ stem cells in a Markov process, with propensities $a_n$ and $b_n$. \textbf{B} The leading eigenvalue of the operator $\mathcal{M}$, $\lambda_1$,  vanishes as the deterministic equation's steady state solution, $\phi^*$, increases (blue line, inset for logarithmic scale). Correspondingly, the expected extinction time, $|\lambda_1|^{-1}$, is large for high $\phi^*$, indicating the presence of a quasi-steady state. The relative spectral gap of the system conditioned on non-extinction, i.e. $|(\lambda_2 - \lambda_1)/\lambda_2|$, increases with $\phi^*$ (orange line). In general, $|\lambda_2|$ is at least $2.6$ times higher than $|\lambda_1|$, supporting a solution of the form of Eq.~(\ref{eq:eigenvectorSolution}) for the vBD master equation. \textbf{C-E} Probability distributions of the vBD model for different $\phi^*$ values and times. All time-dependent PDFs are found by numerically solving the master equation $d {\bf P} / d \tau = \mathcal{M}{\bf P}$ using direct matrix exponentiation. For high $\phi^*$ a quasi-steady state is achieved before relaxation into extinction through transient bimodality (panel \textbf{C}). For intermediate $\phi^*$ transient bimodality is present but a long-lived quasi-steady state is absent (panel \textbf{D}). For low $\phi^*$ extinction is fast via unimodal transient (panel \textbf{E}). In all parameter regimes, the probability distributions of the surviving trajectories reach steady states (insets in \textbf{C-E}). \textbf{F} Behaviours of the vBD model for different $\phi^*$ values and carrying capacities; see main text for explanation of how the phase boundary lines are numerically determined. Note that the discrete jumps of the transition lines reveal the finite-size nature of the niche since the concentrations can only be of the form $n/N$, with $n$ integer, and the carrying capacity $N$ can only adopt integer values.\label{fig:stochasticApproach}}
\end{figure*}

A stochastic treatment of the vBD process is provided by its chemical master equation, i.e., Kolmogorov's forward equation. The chemical master equation (CME) describes the time-evolution of the probability that the system is in one of its states \cite{harris1963theory,gardinerstochastic,schnoerr2017approximation,van1992stochastic}. To construct the CME, we first note that the vBD model only involves reactions that increase or reduce the number of stem cells by one unit. Hence the stochastic process underlying the reaction network (\ref{eq:vBD_CRN}) takes the form of the Markov chain depicted in Fig.~\ref{fig:stochasticApproach}\textbf{A}, where the states $0,1,\dots ,N$ represent the number of stem cells. Note that the vBD process features a reflecting boundary at $n=N$ and an absorbing boundary at $n=0$. The propensities are determined by the law of mass-action and in dimensionless units read: 
\begin{equation}    
    \begin{cases}
        a_n=(n-1)(N-n+1)/[N(1-\phi^*)], \quad n \ge 1\\
        b_n=n+1, \quad n \ge 0.
    \end{cases}
    \label{eq:propensities}
\end{equation} 
Let $P(n,\tau \mid P(\tau_0)=P_0)$ be the probability of finding the system in a state of $n$ cells at time $\tau$, given that it was $P_0$ at time $\tau_0$, which we will abbreviate as $P(n,\tau)$. Defining the probability vector ${\bf P} (\tau)=(P(0,\tau),P(1,\tau),\dots,P(N,\tau))^\mathsf{T}$, where $\mathsf{T}$ denotes the vector transpose, the CME can be expressed as $d {\bf P} / d \tau = \mathcal{M} {\bf P}$, where $\mathcal{M}$ is the operator defined by 
\begin{equation}
    \mathcal{M}=\begin{bmatrix}
        0 & b_0 & 0 & 0 & \ldots \\
        0 & -a_2-b_0 & b_1 & 0 & \ldots \\
        0 & a_2 & -a_3 -b_1 & b_2 \ldots \\
        0 & 0 & a_3 & -a_4 - b_2 &  \ldots \\
        \vdots & \vdots & \vdots & \vdots & \ddots 
    \end{bmatrix}.
\end{equation}

\noindent The $n^\textrm{th}$ row of the master equation reads 
\begin{equation}
\frac{d P}{d \tau} = a_n P(n-1,\tau) + b_n P (n+1,\tau) - (a_{n+1}+b_{n-1}) P(n,\tau), \label{eq:MasterEquation}   
\end{equation}
with $a_0 = b_{-1} = 0$. Note that the only parameters present are the carrying capacity, $N$, and the steady state from the deterministic equations, $\phi^*$, as per \eqref{eq:propensities}. 

The solution of the master equation, for an initial probability distribution ${\bf P}(0)$, is given by ${\bf P} (\tau)= e^{\mathcal{M} \tau} {\bf P} (0)$. The main properties of the solution are captured by the eigenvectors and eigenvalues of $\mathcal{M}$. It is easy to prove that $\lambda_0=0$ is always an eigenvalue associated with the eigenvector $[1,0,\dots,0]^\mathsf{T}$ (the extinction state), while the other eigenvalues are real and negative \cite{smith2015general}. Therefore, the extinction state is always reached, irrespective of the parameter values and initial conditions. Moreover, the expected extinction time is the inverse of the spectral gap, $|\lambda_1 - \lambda_0|^{-1}$, where $\lambda_1$ is the smallest (in absolute value) non-zero eigenvalue. The third eigenvalue, $\lambda_2$, has a considerably higher absolute value than $\lambda_1$, as we can appreciate from the spectral gap of the reduced system obtained by eliminating the extinction state (orange line in Fig.~\ref{fig:stochasticApproach} \textbf{B}) -- we observe that the smallest gap is $\lambda_2 \approx 2.6\lambda_1$ which is achieved in the limit of small $\phi^*$. Hence, after an initial transient the PDF is dominated by the eigenvectors associated with $\lambda_0$ and $\lambda_1$, leading to  
\begin{equation}
    P(n,\tau) \approx {\bf e_0} + e^{-\lambda_1 \tau} {\bf e_1},  
    \label{eq:eigenvectorSolution}
\end{equation}  
\noindent where ${\bf e_0}=[1,0,\dots,0]^\mathsf{T}$ is the extinction state, and ${\bf e_1}=[-1,f(1),f(2),\dots,f(N)]^\mathsf{T}$ is the leading eigenvector. The first element of $\bf{e_1}$ comes from the lower boundary of the state space. The first row in the CME reads $d P(0,\tau) /d\tau =b_0 P(1,\tau)=b_0f(1)e^{-\lambda_1 \tau}$, which leads to (i) $\lambda_1=b_0 f(1)$, and (ii) $P(0,\tau)=1-e^{-\lambda_1 \tau}$. Note that $f(n)$ is the PDF of stem cell numbers conditioned on non-extinction. It follows that the expected extinction time is  
\begin{equation}
    \mathbb{T}=[b_0 f(1)]^{-1}.
\label{eq:extinctionTime1}
\end{equation}

The distribution of the surviving trajectories is defined as
\begin{equation}
    \tilde{\textbf{P}}(\tau)=[\tilde{P_1}(\tau), \dots , \tilde{P_N}(\tau)]^\mathsf{T} = \frac{\textbf{P'}(\tau)}{\sum_{k=1}^{N} P_k (\tau)} = \frac{\textbf{P'}(\tau)}{1-P_0 (\tau)}, 
\end{equation}
where $\textbf{P'}(\tau)=[P(1,\tau), \dots ,P(N,\tau)]^\mathsf{T}$. Hence, the master equation for the PDF conditioned on non-extinction reads
\begin{equation}
    \frac{\partial \tilde{\textbf{P}}}{\partial \tau} = \frac{\partial \textbf{P'}}{\partial \tau} \frac{1}{1-P_0} + \frac{\textbf{P'}}{(1-P_0)^2} \frac{\partial P_0}{\partial \tau}= \tilde{\mathcal{M}}\tilde{\textbf{P}} + b_0 \tilde{P}_1 \tilde{\textbf{P}},  
    \label{eq:MEconditionedonnonextinction}
\end{equation}
where the operator $\tilde{\mathcal{M}}$ results from eliminating the first row and column of $\mathcal{M}$.  Note that, even though the steady state in the original CME is the extinction state, the CME for the surviving trajectories presents a non-trivial steady state (Fig.~\ref{fig:stochasticApproach}{\bf C-E}). In effect, the steady state of Eq.~(\ref{eq:MEconditionedonnonextinction}) yields the entries $f(k)$ of $\bf{e_1}$.   

Numerical experiments reveal the presence of three different behaviours in the solution given by Eq.~(\ref{eq:eigenvectorSolution}). For high $\phi^*$ values the system relaxes to a long-lived quasi-steady state, $f(n)$, before slowly relaxing again to extinction through a transient bimodal distribution (Fig.~\ref{fig:stochasticApproach}{\bf C}). The two modes given by the eigenstates ${\bf e_0}$ and ${\bf e_1}$ in Eq.~(\ref{eq:eigenvectorSolution}) are located at the extinction state and near (but not necessarily at) $N\phi^*$, respectively. The leading eigenvalue vanishes for high $\phi^*$ (blue line in Fig.~\ref{fig:stochasticApproach}{\bf B}), indicating the presence of a quasi-steady state (QSS). For intermediate $\phi^*$ values, the transient bimodality is still present, but the relaxation to extinction is faster than the time to reach the QSS (Fig.~\ref{fig:stochasticApproach}{\bf D}). For low $\phi^*$ values the system rapidly goes extinct, and the mode around $N\phi^*$ is absent (Fig \ref{fig:stochasticApproach}{\bf{E}}). The phase diagram in Fig.~\ref{fig:stochasticApproach}{\bf F} summarises the three types of behaviour of the vBD model as a function of $\phi^*$ and the carrying capacity $N$. To determine the parameter regimes for these three behaviours, we compute numerically the solution to the master equation, ${\bf P} (\tau)= e^{\mathcal{M} \tau}{\bf P} (0)$, choosing $P(n,0)=\delta_{nk}$, where $\delta_{ij}$ is the Kronecker delta and $k=\lceil N\phi^* \rceil$. We then extract the PDF conditioned on non-extinction after the initial transient. The black region in Fig.~\ref{fig:stochasticApproach}{\bf F} corresponds to the case in which such distribution does not present maximum for $n>1$. To distinguish between the bimodal extinction and QSS regions, we compute numerically the second eigenvalue of $\mathcal{M}$, $\lambda_1$, and set a tolerance $\alpha=1\times10^{-2}$. The grey region in Fig.~\ref{fig:stochasticApproach}{\bf F} corresponds to $\lambda_1>\alpha$, and the white region to $\lambda_1\leq \alpha$. In the next two sections we derive approximate solutions for $f(n)$ and $\mathbb{T}$ in the three different regimes.        

\section{Approximate solutions \label{sec:Approximate solutions}}

Here, we present approximate solutions for the vBD master equation. From Eq.~(\ref{eq:eigenvectorSolution}), it follows that the solution adopts the form $P(n,\tau)=f(n)e^{-\lambda_1 \tau}$, $\forall n \geq 1$, and $P(0,\tau)=1-e^{-\lambda_1 \tau}$. Hence, the asymptotic solution is determined by the PDF conditioned on non-extinction, $f(n)$, and the inverse of the expected extinction time, $\lambda_1$. In this section we tackle the problem of approximating $f(n)$, assuming $\lambda_1$ is known, while in the next section we derive an accurate expression for the expected extinction time and correspondingly for $\lambda_1$. 

Substituting Eq.~(\ref{eq:eigenvectorSolution}) into the master equation we obtain 
\begin{equation}
    a_n f(n-1)+b_n f(n+1) +(\lambda_1  - a_{n+1} - b_{n-1})f(n)=0,
\end{equation}
where $1 \leq n < N$. Defining $u_n=f(n)/f(n-1)$ leads to the recurrence relations
\begin{equation}
    u_k=\frac{a_k}{b_{k-1}+a_{k+1}-\lambda_1 - b_k u_{k+1}}, \qquad k \geq 2,
    \label{eq:recurreenceRelationships}
\end{equation}
with the boundary conditions $u_2=(a_2 + b_0 - \lambda_1)/b_1$ and $u_N=a_N/(b_{N-1}-\lambda_1)$. Eq.~(\ref{eq:recurreenceRelationships}) can be solved iteratively, leading to a solution in terms of the continued fraction
\begin{equation}
    u_k= \cFrac{a_k}{c_k} - \cFrac{b_k a_{k+1}}{c_{k+1}} \dots - \cFrac{b_{N-2} a_{N-1}}{c_{N-1}} - \cFrac{b_{N-1} a_{N}}{b_{N-1} - \lambda_1},
    \label{eq:continuedFractionSolution}
\end{equation}
\noindent where we have defined $c_k=\lambda_1 - b_{k-1} - a_{k+1}$. To obtain an expression for $f(n)$, we apply the boundary condition $f(1)=\lambda_1/b_0$, thus leading to 
\begin{equation}
    f(n)=\frac{\lambda_1}{b_0} \prod_{k=2}^{n} u_k.
    \label{eq:generalSolutionf(n)}
\end{equation}
Eqs.~(\ref{eq:continuedFractionSolution}) and (\ref{eq:generalSolutionf(n)}) are exact solutions for the post-transient dynamics given by Eq.~(\ref{eq:eigenvectorSolution}). However, the slow convergence of the continued fraction and the difficulty in applying truncation methods render Eqs.~(\ref{eq:continuedFractionSolution}) and (\ref{eq:generalSolutionf(n)}) unsuitable for the analysis of the system's dynamics. On the other hand, the continued fraction solution offers a fast computational estimation for the master equation's time-dependent solution, often less demanding than direct matrix exponentiation to estimate ${\bf P}(n,\tau)= e^{\mathcal{M}\tau}{\bf P}(0)$. In what follows, we derive three different approximate solutions to the master equation of the vBD process, one for each region of the phase diagram shown in Fig.~\ref{fig:stochasticApproach}{\bf F}. 

\subsection{\label{subsec:QSS}QSS approximation}

For high $\phi^*$ values, $\lambda_1\ll1$ and correspondingly the mean extinction time is very large (see Fig.~\ref{fig:meanFieldApproach}\textbf{B}). Since $\lambda_1=b_0 f(1)$, we can impose a quasi-steady state condition by disregarding the second term in the r.h.s.\ of Eq.~(\ref{eq:MEconditionedonnonextinction}), leading to $\partial \tilde{{\bf P}} / \partial \tau \approx \tilde{\mathcal{M}}\tilde{{\bf P}}$. The steady state of the CME conditioned on non-extinction can be obtained by solving $\tilde{\mathcal{M}}\tilde{\bf{P}}=0$; the $N^\textrm{th}$ row yielding $f(N-1)=(b_{N-1}/a_N)f(N)$. This relationship can be iterated to obtain 
\begin{equation}
    f(k)=\prod_{i=k}^{N-1} \frac{b_i}{\hat{a}_{i+1}}f(N), \qquad 1 \leq k < N
    \label{eq:QSSGardinerGeneral}
\end{equation}
\noindent where $\hat{a}_i=a_i + \delta_{i2} b_0$, and $\delta_{i2}$ is the Kronecker delta. Finally to find $f(N)$ we make use of the normalisation condition which leads us to a closed form solution for the quasi-steady-state PDF
\begin{equation}
    f(N) = \left[ 1 + \sum_{k=1}^{N-1} \prod_{i=k}^{N-1} \frac{b_i}{\hat{a}_{i+1}}\right]^{-1}.  
\end{equation}
 
Substituting the propensities $a_n$ and $b_n$ from Eq.~(\ref{eq:propensities}) yields the approximate QSS solution
 \begin{equation}
     f(k)=\frac{(\Sigma^2)^{N-k} \left[1+\delta_{k1}\frac{(N-1)}{N-1+\Sigma^2}\right]/(N-k)!}{ 1+\sum_{i=1}^{N-1} \frac{(\Sigma^2)^{N-i}}{(N-i)!} \left[1+\delta_{i1}\frac{(N-1))}{N-1+\Sigma^2} \right] },
     \label{eq:QSSvBD}
 \end{equation}
 \noindent for $1 \leq k \leq N$, where we defined $\Sigma^2=N(1-\phi^*)$. The QSS approximation accurately describes the PDF conditioned on non-extinction for high $\phi^*$ values (Figs.~\ref{fig:Fig3}{\bf A} and \ref{fig:Fig4}{\bf A}). 

From the QSS approximation we can calculate the position for the non-zero mode. To do so, it is convenient to study the discrete first derivative $\frac{\partial f(n)}{\partial n}=f(n+1)-f(n)$. For a maximum or minimum to take place, it is necessary that $f(n+1)/f(n) = 1$. According to Eq.~(\ref{eq:QSSvBD}), we have
 \begin{equation}
    \frac{f(n+1)}{f(n)} = \frac{n (N-n)}{N(1-\phi^*)(n+1)} \qquad \forall \; n \geq 2.
\end{equation}
\noindent Solving for $n$ we have 
\begin{equation}
    n^{*\pm} = \frac{1}{2}\left[ N\phi^* \pm \sqrt{(N\phi^*)^2 -4 N (1-\phi^*)} \right].
    \label{eq:QSSmodes}
\end{equation}
\noindent The sign of the discrete second derivative, $ \partial^2 f / \partial n^2 = f(n+1) + f(n-1) -2 f(n)$ reveals that, when both solutions exist, $n^{*+}$ corresponds to a maximum and $n^{*-}$ to a minimum. Note that, for $n^{*\pm}$ to adopt real values, $(N\phi^*)^2 \geq 4 N(1-\phi^*)$, which is the case for all parameter sets in the QSS region. Hence, in general the position of the non-zero mode differs from the deterministic model's prediction, i.e., $n^{*+} \ne N\phi^*$. As an example, for a carrying capacity of $N=100$ and $\phi^*=0.4$, then $n^{*+}\approx 38$. Note that, when $\phi^* \rightarrow 1$ or $N \rightarrow \infty$, $n^{*+} \rightarrow N \phi^*$ and $n^{*-} \rightarrow 0$. Thus in these limits, the distribution has a single mode sitting at the deterministic model's prediction of the mean stem cell number. 


\subsection{\label{sec:SSE}Renormalized system size expansion}

For parameter sets $(\phi^*,N)$ in the bimodal and unimodal extinction regions (grey and black areas in Fig.~\ref{fig:stochasticApproach}{\bf F}), the QSS approximation is unable to capture the PDF conditioned on non-extinction (as it can be seen in Figs.~\ref{fig:Fig3}{\bf B, C} and \ref{fig:Fig4}{\bf A}). In such cases, the condition  $\lambda_1 \ll 1$ does not hold, and hence the second term in the r.h.s.\ of Eq.~(\ref{eq:MEconditionedonnonextinction}) can no longer be considered negligible (which is needed to find the steady state of the CME conditioned on non-extinction iteratively, as we did in the QSS case). Hence a different approximation is needed. In what follows we derive a new approximate PDF conditioned on non-extinction based on a high-order renormalized system size expansion of the vBD's master equation. In section \ref{subsec:RenSSE} we provide a general analytical recipe to obtain an expression for the PDF conditioned on non-extinction, Eqs.~\eqref{eq:discrete_LNA} and \eqref{Eq:sseOrderNsolution} being the main results. While self-contained, this derivation is mathematically lengthy, and the reader can skip it should they be solely interested in its application to the vBD process. We point the reader to \cite{thomas2015approximate} for a detailed derivation of the method. In section \ref{subsec:SSE_vBD} we apply the renormalized SSE results to the vBD system, and obtain an analytical expression for the PDF conditioned on non-extinction. 

\subsubsection{Derivation of the renormalized SSE formula \label{subsec:RenSSE}}

The van Kampen system size expansion (SSE) approximates a master equation by splitting the random variable that describes the stem cell number, $n$, into deterministic and non-deterministic components, to then obtain a master equation for the non-deterministic components (usually taken as a continuous random variable). The resultant CME can be expanded in powers of $N^{-1/2}$, truncated to a desired order and solved, leading to a hierarchy of approximate solutions \cite{van1976expansion}. Whilst the bulk of applications are centred in truncating the CME after the order $N^0$ to obtain the linear noise approximation (LNA), which considers the particle concentration equal to its deterministic value, other works have used higher order truncation schemes to obtain corrections to the mean concentration \cite{grima2009noise,grima2010effective}. The deterministic component is commonly assumed to be given by the solution of the rate equations. However, the disagreement between the deterministic and stochastic predictions (see Fig.~\ref{fig:meanFieldApproach}{\bf B}) makes it sensible to introduce a correction term to the mean stem cell concentration in the original ansatz. To do so, we follow the procedure from \cite{thomas2015approximate}, that starts by considering the ansatz 
\begin{equation}
    \frac{n}{N}=\phi + N^{-1/2} \langle \epsilon \rangle + N^{-1/2} \hat{\epsilon},
    \label{eq:vanKampenAnsatzRenormalized}
\end{equation}
where the first term in the r.h.s.\ is the zero-order mean stem cell concentration obtained from the deterministic rate equations, the second term is a correction to the mean concentration due to fluctuations, and the third term in the r.h.s.\ represents fluctuations about the corrected mean concentration. This renormalization of the mean concentration leads to a different system-size expansion than the conventional one by van Kampen, which we refer to as the renormalized system-size expansion. 

Next, we briefly describe how to compute the corrections to the mean concentration as a series in powers of $N^{-1/2}$:
\begin{equation}
    \langle \epsilon \rangle = \sum_{j=0}^{\infty} N^{-j/2} a_1^{(j)}.\\
    \label{eq:expansionMean}
\end{equation}
The expansion coefficients for the correction term to the mean concentration are calculated iteratively as follows:
\begin{equation}
\begin{multlined}
    a_n^{(j)}= - \frac{1}{n \mathcal{J}} \sum_{k=1}^j \sum_{s=0}^{\lceil k/2 \rceil} \sum_{p=1}^{k-2(s-1)}
    \mathcal{D}_{p,s}^{k-p-2(s-1)} \sum_{m=0}^{3(j-k)} a_m^{(j-k)} \mathcal{I}_{mn}^{p,k-p-2(s-1)},
    \label{eq:SSEexpCoefNORen}
\end{multlined}
\end{equation}
where $\mathcal{J}$ is the Jacobian of the deterministic rate equations, and we assume $a_{m}^{(0)}=0$. To define the operators $\mathcal{D}^q_{p,s}$, we assume that the propensity functions for a birth or death event when the system is in a state of $n$ stem cells ($a_{n+1}$ and $b_{n-1}$), expressed in terms of the concentrations, can be expanded in power series of the inverse carrying capacity ($N^{-1}$) as
\begin{align}
a(N \phi,N)= N \sum_{s=0}^{\infty} N^{-s} g_1^{(s)}(\phi),\\
b(N \phi,N)= N \sum_{s=0}^{\infty} N^{-s} g_2^{(s)}(\phi).
\label{eq:expansionOfPropensities}
\end{align}
where $g_r^{(s)}(\phi)$ are the expansion coefficients. For example, $g_1^{(s)}(\phi)$ can be obtained by defining $z=N^{-1}$, transforming $a(N \phi, N)/N \rightarrow za(\phi/z,1/z)$, and Taylor-expanding around $z=0$. Following we define 
\begin{equation}
 \mathcal{D}_{p,s}^q= \sum_{r=
 1}^2 (S_r)^p \frac{\partial^q g_r^{(s)}(\phi)}{\partial \phi^q},   
\end{equation}
where $S_r$ is the net change in the number of stem cells when the $r^{\textrm{th}}$ reaction occurs, namely $S_1=1$ and $S_2=-1$. The functions $\mathcal{I}_{mn}^{\alpha \beta}$ in Eq.~(\ref{eq:SSEexpCoefNORen}) are defined as
\begin{equation}
\begin{multlined}
\mathcal{I}_{mn}^{\alpha \beta} = \frac{\sigma^{\beta -\alpha + n - m}}{\alpha !} \sum_{s=0}^{min(n-\alpha, m)} \binom{m}{s} 
 \frac{[\beta + \alpha + 2s - (m+n) -1]!!}{[\beta + \alpha + 2s - (m+n)]! (n-\alpha -s)!},
\end{multlined}
\end{equation}
for $(\alpha+\beta) - (m+n)$ even, and zero otherwise. We have introduced the notation $(2k-1)!!=(2k)! / (2^k k!)$ for the double factorial, and  $\sigma$ is the standard deviation of the concentration according to the standard linear noise approximation \cite{van1976expansion}
\begin{equation}
    \frac{\partial \sigma^2}{\partial \tau}= 2 \mathcal{J} \sigma^2 + \mathcal{D}_{0,0}^2.
    \label{eq:eq25}
\end{equation}

The variance of the fluctuations can also be computed as a series in powers of $N^{-1/2}$
\begin{equation}
    \hat{\sigma}^2=\sigma^2 + \sum_{j=1}^{\infty} N^{-j/2} \hat{\sigma}_j^2, 
    \label{eq:expansionVariance}
\end{equation}
where the expansion coefficients are given by
\begin{equation}
     \hat{\sigma_{j}^2} = 2 \left[ a_2^{(j)} - \mathcal{B}_{j,2}\left(\left\{\chi! a_1^{(\chi)}\right\}^{j-1}_{\chi=1}\right)/j! \right],
\label{eq:varianceExpansionCoef}
\end{equation}
with $\mathcal{B}_{j,k}$ being the partial Bell polynomials, where $\{\cdot \}$ denotes the set of arguments \cite{andrews1998theory}. For example, $\mathcal{B}_{4,2}\left(\left\{ \chi! a_1^{(\chi)} \right\} \right)$ has as arguments $1!a_1^{(1)}$, $2!a_1^{(2)}$, and $3!a_1^{(3)}$.

To summarise, the system size expansion procedure involves expanding the master equation for the fluctuations about the mean concentration (assuming the fluctuations are a continuous random variable) in powers of $N^{-1/2}$, and truncating after $\mathcal{O}(N^0)$ to obtain a Fokker-Planck equation that yields the LNA, thus describing Gaussian fluctuations around the mean concentration. The higher order approximate solutions can then be expressed in terms of the first-order approximation. However, this approach often leads to non-physically meaningful distributions for truncations of the system-size expansion beyond the LNA level of approximation, e.g., yielding negative probabilities or oscillatory behaviour. Such effects can be greatly reduced by introducing a discrete formulation of the SSE's approximate solutions \cite{thomas2015approximate}.

The discrete formulation replaces the continuous-variable LNA approximate solution, by the discrete approximation
 \begin{equation}
    P_0(n,\tau)= \frac{1}{2} \frac{e^{-\frac{x^2}{2 \Sigma^2}}}{\sqrt{2 \pi} \Sigma} \left[ \text{erf}\left(\frac{ix + \pi \Sigma^2}{\sqrt{2} \Sigma} \right) - \text{erf}\left(\frac{ix - \pi \Sigma^2}{\sqrt{2} \Sigma} \right) \right],
    \label{eq:discrete_LNA}
 \end{equation}
 where erf is the error function, $x=n-N\phi-N^{1/2} \langle \epsilon \rangle$ is the stem cell number centred about its (corrected) deterministic value, and $\Sigma^2= N \hat{\sigma}^2$ its variance. Note that the time dependence is implicit in the temporal change of the mean concentration and the variance of concentration fluctuations, $\phi(\tau)$ and $\sigma(\tau)$, respectively. Equation \eqref{eq:discrete_LNA} is the discrete version of a Gaussian in the sense that every moment of the distribution coincides with their corresponding of a continuous-variable Gaussian. Next, the expansion of the SSE approximate solution up to any order can be expressed in terms of $P_0(n,\tau)$ and its derivatives 
 \begin{equation}
        P(n,\tau)= P_0 (n,\tau) + \sum_{j=1}^\infty N^{-\frac{j}{2}}
        \sum_{m=1}^{3j}  \hat{a}_m^{(j)} (-N^{\frac{1}{2}} \partial_n)^m P_0 (n,\tau)
    \label{Eq:sseOrderNsolution}
 \end{equation}
\noindent where the new expansion coefficients $\hat{a}_m^{(j)}$ are related to the coefficients found earlier ($a_m^{(j)}$ in Eq.~(\ref{eq:SSEexpCoefNORen})) by the relation
\begin{equation}
    \hat{a}_m^{(j)}=\sum_{k=0}^j \sum_{n=0}^{3k} a_n^{(k)} k_{m-n}^{(j-k)},
\label{eq:sse_coeff_ren}
\end{equation}
with
\begin{equation}
\begin{multlined}
    k_j^{(n)}=\frac{1}{n!}\sum_{m=0}^{\lfloor j/2 \rfloor} (-1)^{j+m} \sum_{k=j-2m}^{n-m} \binom{n}{k} \\ \times \mathcal{B}_{k,j-2m}\left(\left\{\chi! a_1^{(\chi)}\right\}_{\chi=1}^{k-j+2m+1}\right) \\
    \times
    \mathcal{B}_{n-k,m}\left(\left\{\frac{\chi!}{2} \hat{\sigma}^2_{(\chi)}\right\}_{\chi=1}^{n-k-m+1}\right).
\end{multlined}
\end{equation}
To calculate the derivatives of $P_0(n,\tau)$, it is possible to prove by induction the following formula
\begin{equation}
    \begin{multlined}
        \frac{\partial^m P_0}{\partial y^m} = - \frac{(m-1)}{\Sigma^2} \frac{\partial^{m-2}P_0}{\partial y^{m-2}} - \frac{y}{\Sigma^2} \frac{\partial^{m-1} P_0}{\partial y^{m-1}} 
        + (-1)^{\lceil m/2 \rceil +1} F(y,m) \pi^{m-2} \frac{e^{- \frac{\pi^2 \Sigma^2}{2}}}{\Sigma^2},    
    \end{multlined}
    \label{eq:P0_derivatives}
 \end{equation}
where $\lceil \cdot \rceil$ denotes the ceiling function, and $F(y,m)$ is $\sin(\pi y)$ for $m$ odd and $\cos(\pi y)$ for $m$ even. 

Calculation of the correction terms to the mean concentration and the variance of fluctuations, $\langle \epsilon \rangle$ and $\hat{\sigma}^2$ respectively, by truncation of Eqs.~\eqref{eq:expansionMean} and \eqref{eq:expansionVariance} to any desired order, followed by substitution into Eq.~(\ref{Eq:sseOrderNsolution}) and truncation, provides a means to systematically obtain approximate solutions to the CME. Note that the order of the approximate solution is determined by the order of truncation of $\langle \epsilon \rangle$ and $\hat{\sigma}^2$. For example, to obtain an approximation of order $(s+1)/2$ with $s=0,1,\dots$, both $\langle \epsilon \rangle$ and $\hat{\sigma}^2$ are expanded up to that order (assuming that at least one of the coefficients is non-zero) using Eqs.~\eqref{eq:expansionMean} and \eqref{eq:expansionVariance}, and then the approximate distribution is obtained by truncating Eq.~\eqref{Eq:sseOrderNsolution} after the order $(s+1)/2$.

\subsubsection{The expansion of the vBD master equation \label{subsec:SSE_vBD}}

We now employ the renormalized discrete-formulation of the system size expansion to approximate the probability distributions conditioned on non-extinction in the bimodal extinction region, i.e., the grey zone in Fig.~\ref{fig:stochasticApproach}{\bf F}. To do so, we observe that the absorbing boundary, which is the factor that renders the deterministic rate equations inaccurate for calculating the evolution of the mean stem cell number, is absent when conditioning on non-extinction. In effect, the rate equations capture meaningful information about the mean stem cell number of the surviving stochastic trajectories. Thus, to approximate the steady states of Eq.~(\ref{eq:MEconditionedonnonextinction}) we can apply the
renormalized system size expansion described in the previous subsection, under stationary conditions, to the non-conditional master equation (\ref{eq:MasterEquation}), which includes the extinction state. We then recover the PDF conditioned on non-extinction by removing the extinction state and multiplying the resulting distribution by a normalization constant $1/(1-P(0))$, where $P(0)$ is the probability of being in the extinction state obtained by the SSE. In particular, to capture non-Gaussian fluctuations, we truncate the renormalized SSE after terms of $\mathcal{O}(N^{-1})$. 

The expansion coefficients for the vBD propensities ($g_r^{(s)}$ in Eq.~\eqref{eq:expansionOfPropensities}) are given by
\begin{equation}
g_1^{(0)}(\phi)=\phi(1-\phi)/(1-\phi^*); \quad g_2^{(0)}(\phi)=\phi,     
\end{equation}
and $g_r^{(s)}(\phi)=0 \; \forall s \geq 1$, which allows us to calculate the values of $D_{p,s}^q$.

From the standard LNA we obtain the order $N^0$ approximation for the first two moments of the distributions
\begin{align}
       \frac{\partial \phi}{\partial \tau} &= \frac{(\phi^*-\phi)\phi}{1-\phi^*},  \\
        \frac{\partial \sigma^2}{\partial \tau} &=\frac{1}{1-\phi^*}[2(\phi^*-2\phi)\sigma^2 + \phi(2-\phi-\phi^*)],
    \label{eq:varianceODE}
\end{align}
where the first line is the rate equation and the second one is Eq.~(\ref{eq:eq25}) in the vBD case. These equations predict the first two moments to be $\phi=\phi^*$ and $\sigma^2=1-\phi^*$ under stationary conditions. The Jacobian in stationary conditions reads $\mathcal{J}=- \phi^* /(1-\phi^*)$. We can now calculate the first few expansion coefficients $a_n^{(j)}$ from Eq.~(\ref{eq:SSEexpCoefNORen}):
\begin{equation}
    \begin{aligned}
    a_1^{(1)} &= \phi^*-1,\\
    a_3^{(1)} &= \frac{(1-\phi^*)^2}{6\phi^*}\left[(2-3\phi^*)(1-\phi^*)-2\right],\\
    a_2^{(2)} &= \frac{1}{4\phi^*} \left[ a_1^{(1)} (3\phi^*-4) - 12a_3^{(1)} -(1-\phi^*) \right], \\
    a_4^{(2)} &= \frac{(1-\phi^*)}{8\phi^*}\left[ a_3^{(1)} \frac{(5\phi^*-8)}{1-\phi^*} - a_1^{(1)} \phi^* - \phi^*/6 -1 \right],\\
    a_6^{(2)} &= \frac{1}{2} \left(a_3^{(1)}\right)^2,
\label{eq:anj_vBD_nonren}    
    \end{aligned}
\end{equation}
whilst $a_n^{(j)}=0$ for $n+j$ odd. 
Next, we expand the corrections to the mean concentration and fluctuation's variance ($\langle \epsilon \rangle$ and $\hat{\sigma}^2$ in Eqs.~\eqref{eq:expansionMean} and \eqref{eq:expansionVariance}) up to order $N^{-1}$: $\langle \epsilon \rangle \approx N^{-1/2} a_1^{(1)} + N^{-1} a_1^{(2)}$, and $\hat{\sigma}^2 \approx\sigma^2 + N^{-1/2} \hat{\sigma}_1^2 + N^{-1} \hat{\sigma}_2^2$. Eq.~(\ref{eq:varianceExpansionCoef}) yields $\hat{\sigma}_1^2=0$ and $\hat{\sigma}_2^2=2\left[a_2^{(2)}-\left(a_1^{(1)}\right)^2\right]$. Thus, we arrive at
\begin{equation}
\begin{aligned}
    \langle \epsilon \rangle &\approx N^{-1/2} a_1^{(1)},\\
    \hat{\sigma}^2 &\approx 2 N^{-1} \left[a_2^{(2)}-\left(a_1^{(1)}\right)^2\right],
\end{aligned}
\end{equation}
which, upon substitution on Eq.~(\ref{eq:discrete_LNA}) leads to our order $N^0$ (LNA) approximation:
\begin{equation}
    f_0(n)=\frac{P_0(n)}{1-P_0(0)}.
\end{equation}
Note that, since we have imposed the stationary conditions $\phi=\phi^*$ and $\sigma=1-\phi^*$, there is no longer a time dependence in $P_0$. To calculate the higher order approximate solutions we make use of Eq.~(\ref{Eq:sseOrderNsolution}). The expansion coefficients, corrected by renormalization, are given by Eq.~(\ref{eq:sse_coeff_ren}). The first few coefficients are
\begin{equation}
    \begin{aligned}
        \hat{a}_1^{(1)} &= \hat{a}_2^{(1)}=\hat{a}_1^{(2)}=\hat{a}_2^{(2)}=\hat{a}_3^{(2)}=\hat{a}_5^{(2)}=0,\\
        \hat{a}_3^{(1)} &= \frac{(1-\phi^*)^2}{6\phi^*}\left[(2-3\phi^*)(1-\phi^*)-2\right],\\
        \hat{a}_4^{(2)} &= \frac{(1-\phi^*)^2}{2\phi^*} \left( 3\phi^* -\frac{5}{4} \right) - \frac{1- \phi^*}{48} -\phi^* 
        \\ & + \frac{(2-\phi^*)(8-5\phi^*)}{48 \phi^* (1-\phi^*)}, \\
        \hat{a}_6^{(2)} &= \frac{1}{72}(1-\phi^*)^2.
    \end{aligned}
\end{equation}
Remarkably, there are fewer non-zero renormalized coefficients than regular ones; hence, the analytical expressions for the SSE distributions corrected by renormalization adopt a simpler form. 

Next, Eq.(~\ref{Eq:sseOrderNsolution}) yields the following expression for the order $N^{-1/2}$ approximate solution 
\begin{equation}
\begin{aligned}
    P_1(n) &= P_0(n)- N^{-1/2} \left\{ \hat{a}_1^{(1)} N^{1/2} \frac{\partial P_0}{\partial n} + \hat{a}_3^{(1)} N^{3/2} \frac{\partial^3 P_0}{\partial n^3} \right\}, \\
    f_1(n) &= \frac{P_1(n)}{1-P_1(0)}.
\end{aligned}
 \end{equation}
Finally the order $N^{-1}$ approximate solution is
\begin{equation}
\begin{aligned}
    P_2(n) &=P_1(n)+ N^{-1} \left\{ \hat{a}_4^{2} N^2 \frac{\partial^4 P_0}{\partial n^4} + \hat{a}_6^{2} N^3 \frac{\partial^6 P_0}{\partial n^6} \right\},\\
    f_2(n) &=\frac{P_2(n)}{1-P_2(0)}.
\end{aligned}
\label{eq:RenSSE}
 \end{equation}
 
\begin{figure*}[htbp]
    \begin{center}
        \includegraphics[scale=0.46]{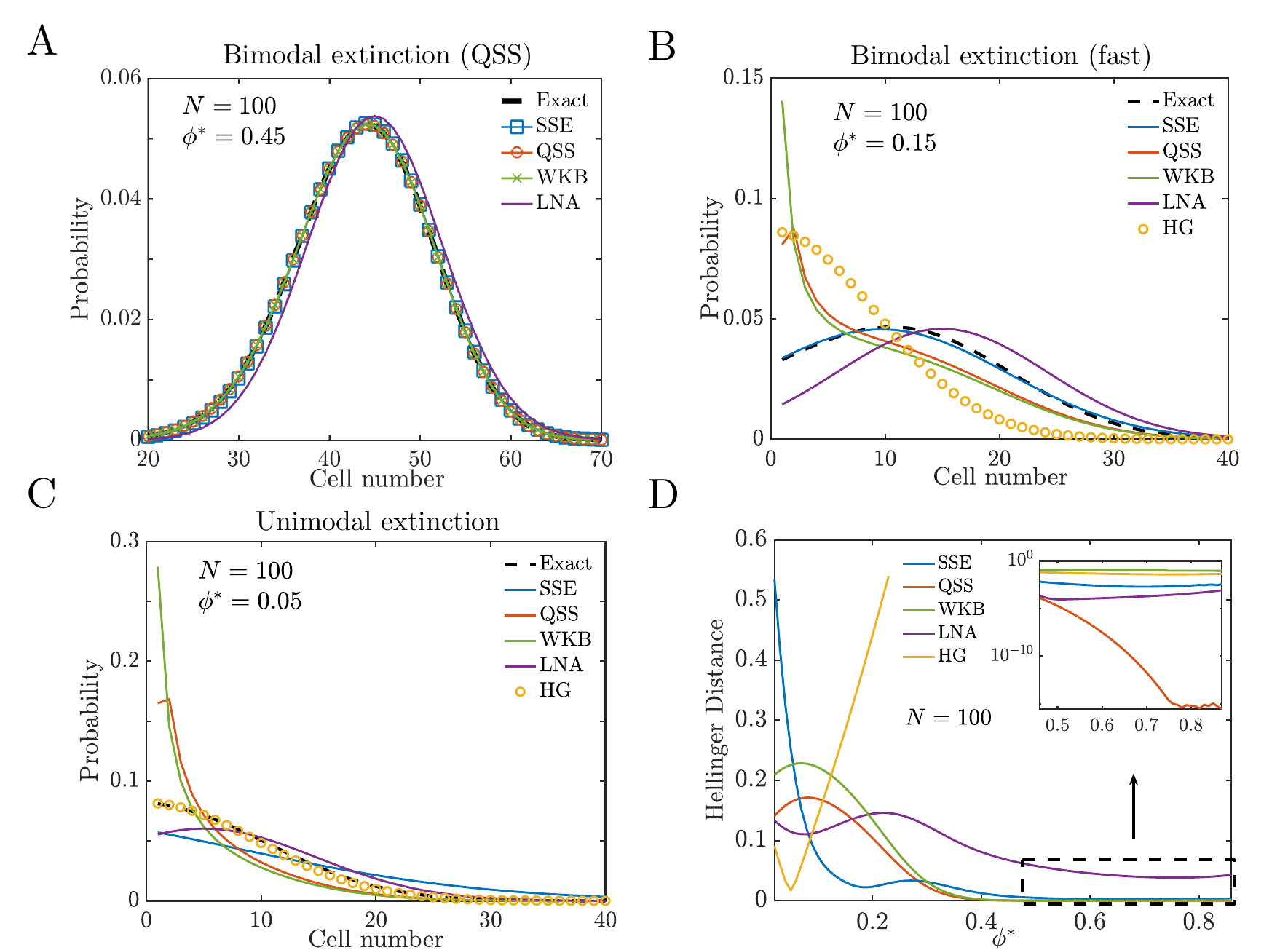}
        \caption{Approximate solutions of the vBD master equation. The SSE, QSS, WKB, LNA, and HG lines correspond to the renormalized system-size expansion (up to order $N^{-1}$), quasi-steady state approximation, WKB approximation, linear-noise, and half Gaussian approximations, respectively. {\bf A} For $(N,\phi^*)$ within the bimodal extinction region the QSS, SSE, and WKB provide good approximations of the probability distribution conditioned on non-extinction, $f(n)$, although the QSS is the most accurate one (see inset in panel {\bf D}). {\bf B} In the bimodal extinction region, the renormalized SSE is the only accurate approximation. {\bf C} In the unimodal extinction region, the HG provides the best approximation. {\bf D} The Hellinger distance between the various distribution approximations and the exact distribution confirms that the QSS, SSE, and HG are the best approximations for the quasi-steady state, bimodal extinction, and unimodal extinction regions, respectively. 
        }   
        \label{fig:Fig3}
    \end{center}
\end{figure*}

\begin{figure*}
    \begin{center}
        \includegraphics[scale=0.46]{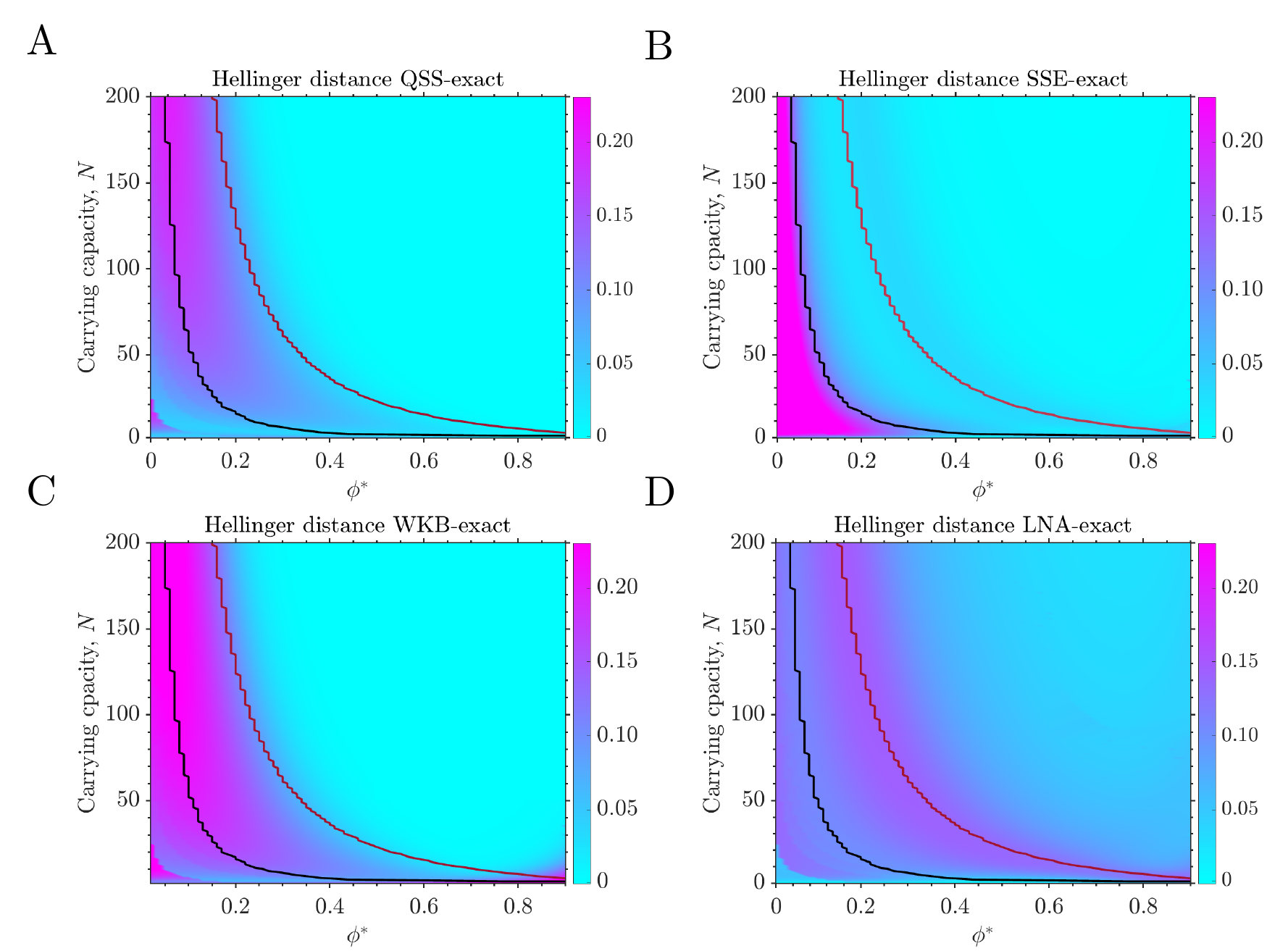}
    \end{center}
    \caption{Hellinger distance between the exact and approximate probability distributions conditioned on non-extinction, as a function of the carrying capacity, $N$, and the deterministic rate equation's steady state, $\phi^*$. Red lines represent the interfaces between the QSS and the non-QSS regions, while the black lines are the interfaces between the transient unimodal and bimodal extinction regions; note these lines demarcate the three different phases shown in Fig.~\ref{fig:stochasticApproach}{\bf F}. For high carrying capacities, the QSS approximation performs well at capturing the PDF conditioned on non-extinction (bimodal extinction region; see panel {\bf A}). The renormalized SSE is accurate in both fast and QSS bimodal extinction regions (see panel {\bf B}). The WKB has a similar range of validity as the QSS approximation (see panel {\bf C}). The LNA is noticeably less accurate than the other three approximations except for low $\phi^*$ (unimodal extinction region; see panel {\bf D}).} 
    \label{fig:Fig4}
\end{figure*}

The renormalized SSE accurately describes the probability conditioned on non-extinction for $\phi^*$ in the QSS region (Figure \ref{fig:Fig3}{\bf A}), as well as in the fast bimodal extinction region (Fig.~\ref{fig:Fig3}{\bf B}), where the QSS approximation breaks down, and accurately captures the distribution skewness. Moreover, this result is robust to changes in the carrying capacity (Fig.~\ref{fig:Fig4}{\bf B}). However, the accuracy of the renormalized SSE decreases dramatically for very low $\phi^*$ values, as it can be seen in Fig.~\ref{fig:Fig3}{\bf C}. For non-linear birth-death processes featuring non-Gaussian fluctuations, the renormalized SSE consistently performs better than the LNA, which is unable to capture the distribution skewness under stationary conditions.

\subsubsection{Approximation for low $\phi^*$ values}

The phases with extinction through a bimodal and unimodal transient are separated by a critical curve, which also accurately demarcates the regions of parameter space where the renormalized SSE expansion is accurate and where it is not (see Fig.~\ref{fig:Fig4}{\bf B}). We attribute the inaccuracy of the SSE in the low $\phi^*$ region to the fact that the PDF conditioned on non-extinction no longer features a mode around $N\phi^*$, which is the basis of the SSE approximations. 

In the unimodal extinction region -- black region in Fig.~\ref{fig:stochasticApproach}{\bf F}-- the time-dependent probability distribution features a mode at $n=0$ as extinction is approached. This suggests that we can approximate the cell number concentration by the steady-state of the deterministic equations, $\phi^*$, and the cell number concentration fluctuations (conditioned on non-extinction) by means of the LNA, Eq.~(\ref{eq:varianceODE}), yielding $\sigma^2=1-\phi^*$ under stationary conditions. Note that the latter conditions naturally arise from the conditioning of the distribution on non-extinction. Given the mode at zero, a Gaussian is clearly not a good approximation and hence instead we try a half-Gaussian approximation with the aforementioned first two moments
\begin{equation}
   f(n) = \sqrt{\frac{2}{N(1-\phi^*) \pi}} e^{- n^2 /[2N(1-\phi^*)]}.
\end{equation}
The half-Gaussian provides an excellent approximation to the PDF conditioned on non-extinction in the low $\phi^*$ unimodal region (Figure \ref{fig:Fig3}  {\bf C}). However, as expected, this approximation breaks down for higher $\phi^*$ values, where the non-trivial mode is present (Fig.~\ref{fig:Fig3}{\bf B} and {\bf D}). 

\subsection{\label{sec:WK} WKB approximation}

An alternative way to obtain a quasi-steady state approximation is the popular Wentzel–Kramers–Brillouin (WKB) approximation \cite{cianci2015wkb,assaf2006spectral,assaf2017wkb}. The WKB approximation often leads to simpler expressions for a QSS than Eq.~(\ref{eq:QSSGardinerGeneral}). A detailed derivation of the WKB approach to solve master equations in quasi-steady state conditions can be found in \cite{cianci2015wkb}, although we repeat the main ideas in what follows.    

The WKB approximation starts by transforming the vBD's PDF of observing $n$ cells, $P(n)$ to a continuous PDF for the cell concentration, $\phi=n/N$, assuming $N$ sufficiently large. The vBD master equation reads   \begin{equation}
    \frac{\partial P}{\partial \tau} = a_n P(n-1,\tau) + b_n P(n+1,\tau) - (a_{n+1}+b_{n-1})P(n,\tau).
\end{equation}
Defining $\phi=n/N$, we can transform the propensities to $\Omega_+ (\phi)=a_{n+1}/N$ and $\Omega_- (\phi)=b_{n-1}/N$, arriving at 
\begin{equation}
        \Omega_+ (\phi) = \frac{\phi (1- \phi)}{1- \phi^*}; \qquad \Omega_- (\phi)= \phi.
\end{equation}
We can now write the master equation for the PDF of the continuous variable $\phi$, $\Pi(\phi,\tau)=P(N\phi,\tau)$. It follows that $P(n\pm1,\tau)=P(N(\phi \pm 1/N),\tau)=\Pi(\phi \pm 1/N,\tau)$. Mutiplying the CME by $1/N$ and applying the quasi-stationary condition $\partial P / \partial \tau = 0$ yields
\begin{equation}
    \begin{multlined}
    \Omega_+\left(\phi - \frac{1}{N}\right) \Pi \left(\phi - \frac{1}{N}\right)  + \Omega_-\left(\phi + \frac{1}{N}\right) \Pi\left(\phi+\frac{1}{N}\right) 
    - \left( \Omega_-(\phi) + \Omega_+ (\phi) \right) \Pi(\phi) =0.
    \end{multlined}
\end{equation}
Next, the WKB approximation amounts to assuming a solution of the form, 
\begin{equation}
    \Pi(\phi)=K(\phi) e^{-N S(\phi)} \left[1+ \mathcal{O}\left( \frac{1}{N} \right) \right],
\end{equation}
\noindent where $S(\phi)$ and $K(\phi)$ are of the order of unity. Substituting in the quasi-stationary master equation, expanding with respect to $N^{-1}$, and collecting the leading order terms yields
\begin{equation}
    \Omega_+(\phi) e^{S'(\phi)-1} + \Omega_-(\phi) e^{-S'(\phi)-1} =0, 
\end{equation}
\noindent where $S'(\phi)=dS(\phi)/d \phi$. From here, we note that the above equation corresponds to a stationary Hamilton-Jacobi equation ($H(\phi,S'(\phi))=0$), for an action $S$ with Hamiltonian
\begin{equation}
H(\phi,p)= \Omega_+(\phi) e^{p-1} + \Omega_-(\phi) e^{-p-1}, 
\label{eq:Hamiltonian}
\end{equation}
with $p=S'(\phi)$. The corresponding Hamilton equations are 
\begin{equation}
    \begin{cases}
    \Dot{\phi}=\frac{\partial H}{\partial p}=\Omega_+(\phi) e^{p} + \Omega_-(\phi) e^{-p}, \\
    \Dot{p}=-\frac{\partial H}{\partial \phi}= (1-e^{p}) \frac{\partial \Omega_+ (\phi)}{\partial \phi} + (1-e^{-p}) \frac{\partial \Omega_- (\phi)}{\partial \phi}.
    \end{cases}
    \label{eq:HamiltonEquations}
\end{equation}
Since we are interested in the zero-energy solution ($H=0$), with initial conditions $\phi(t_0)=\phi^*$, the action along a fluctuation trajectory will be given by 
\begin{equation}
    S=\int_{t_0}^{t} p  \; \Dot{\phi} \; dt'.
    \label{eq:action}
\end{equation}
Hence, we can find $S$ by solving Hamilton's equations and integrating $p \Dot{\phi}$. From Hamilton's equations \eqref{eq:HamiltonEquations}, we note that there is a trivial solution with $p=0$. This solution leads to the deterministic rate equations and hence is of no interest to us. The other solution comes from setting $H=0$ in the Hamiltonian expression \eqref{eq:Hamiltonian}, solving for $p$, and substituting in Hamilton's equations, to yield
\begin{equation}
    p=\log\left(\frac{\Omega_- (\phi)}{\Omega_+ (\phi)} \right); \qquad \Dot{\phi} = \Omega_-(\phi) -  \Omega_+ (\phi). 
\end{equation}
\noindent We are now ready to calculate the action $S(\phi)$. Substituting in Eq.~(\ref{eq:action}) and integrating, we obtain
\begin{equation}
    S(\phi)-S(\phi^*)=(1-\phi) \log \left( \frac{1-\phi}{1-\phi^*} \right) + \phi - \phi^*.
\end{equation}
It is possible to prove, using the next order contributions to the WKB expansion, that the prefactor $K(\phi)$ is
\begin{equation}
    K(\phi)=A \left( \Omega_+(\phi) \Omega_-(\phi)\right)^{-1/2}, 
\end{equation}
where $A$ is later determined by the normalisation condition \cite{cianci2015wkb}. Finally, the WKB expansion for the vBD system yields the probability distribution
\begin{equation}
    \Pi(\phi)=\sqrt{\frac{N}{2\pi (1-\phi)}} \;\; \frac{\phi^*}{\phi} \; e^{N(\phi^*-\phi)} \left( \frac{1-\phi}{1-\phi^*}\right)^{N(\phi -1)}.
\end{equation}
 While the WKB is a good approximation for large values of $\phi^*$ (Figure \ref{fig:Fig3}{\bf A}), it fails for small and intermediate values of $\phi^*$ (\ref{fig:Fig3} {\bf B-C}). In general, the range of validity of the WKB approximation coincides with that of the QSS approximation (compare Figs.~\ref{fig:Fig4}{\bf A} and {\bf C}). An advantage of the WKB approximation stems from its simple analytical form. However the QSS approximation is generally more accurate than the WKB approximation (see Fig.~\ref{fig:Fig3}{\bf D}).  

\section{\label{sec:ExtinctionTime}Calculation of the extinction time}

\begin{figure*} [htbp!]
    \begin{center}
        \includegraphics[scale=0.46]{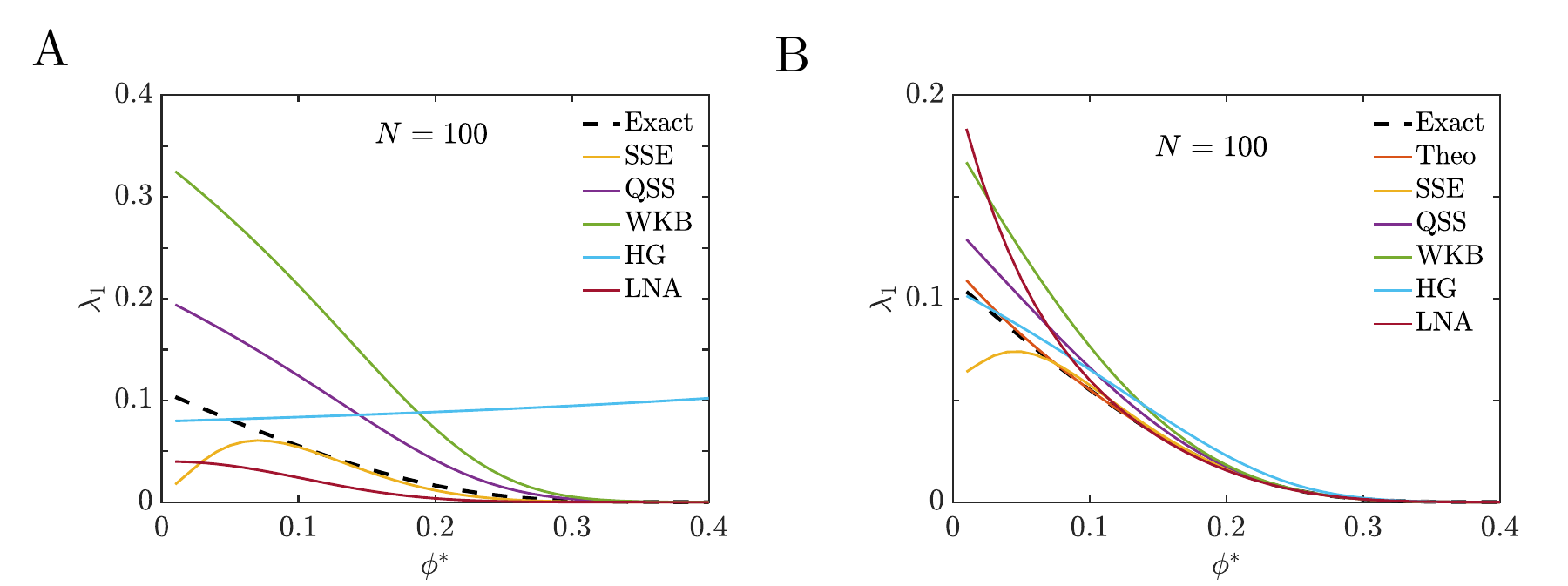}
    \end{center}
    \caption{Inverse of the expected extinction time, $\lambda_1$, as a function of $\phi^*$ for a fixed carrying capacity of $N=100$. \textbf{A} $\lambda_1$ estimated as $b_0 f(1)$, following Eq.~(\ref{eq:extinctionTime1}), with $f(1)$ extracted from the PDF approximations shown in Fig.~\ref{fig:Fig3}. \textbf{B} $\lambda_1$ calculated from Eq.~(\ref{eq:meanExtinctionTime}) and Eq.~(\ref{eq:extinctionTimes}), using different approximations of the probability distribution conditioned on non-extinction. The black dashed line represents the exact $\lambda_1$ value, numerically obtained from the leading eigenvalue of the master operator. The $\lambda_1$ estimation via Eq.~(\ref{eq:meanExtinctionTime}) with $f(n)$ calculated numerically (labelled ``Theo'', orange line) is in good agreement with the exact $\lambda_1$. For very low $\phi^*$, Eq.~(\ref{eq:meanExtinctionTime}) with the half Gaussian approximation is the best fit, while for higher $\phi^*$ the SSE performs better. For high $\phi^*$ all approximations except the half Gaussian lead to good estimates of $\lambda_1$. The application of Eq.~(\ref{eq:meanExtinctionTime}) outperforms the direct approximation of $\lambda_1$ using  Eq.~(\ref{eq:extinctionTime1}) (contrast panels {\bf A} and {\bf B}).}
    \label{fig:extinctionTimes}
\end{figure*}

In the previous section we have shown different approximations for the leading eigenvector of the vBD model's master equation solution, Eq.~(\ref{eq:eigenvectorSolution}). The time-dependent component of the solution is determined by the leading eigenvalue, $\lambda_1$, which is the inverse of the expected extinction time. The most direct method for estimating this eigenvalue would be to calculate the probability conditioned on non-extinction, $f(n)$, and making use of Eq.~(\ref{eq:extinctionTime1}) to yield $\lambda_1=b_0 f(1)$. However, the absorbing boundary at $n=0$ leads to an inaccurate estimation of the probability of having $n=1$ stem cells (using all approximation methods considered) and thus we cannot use this method to estimate the extinction time (see Fig.~\ref{fig:extinctionTimes}{\bf{A}}). Hence, we present an alternative calculation for the expected extinction time that is based on averaging the mean extinction time starting from any state, and makes use of the Kolmogorov's backward equation. This approach has been effectively used for other similar problems (see for example \cite{ashcroft2015statistical,holehouse2021non}). Instead of relying solely on the estimation of $f(1)$, this method involves averaging among all the $f(n)$ values, which significantly improves the accuracy with respect to direct application of $\lambda_1=b_0 f(1)$. 

Given an estimate for the probability distribution of the surviving trajectories, $f(n)$, the expected extinction time $\mathbb{T}$ is simply the average among initial conditions of the mean first passage times to hit the extinction state 
 \begin{equation}
     \mathbb{T}= \sum_{n=1}^{N} \tau^*_n f(n), 
     \label{eq:meanExtinctionTime}
 \end{equation}
where $\tau^*_n$ is the mean first passage time to hit the extinction state, starting from the state with $n$ cells. Hence, to estimate $\mathbb{T}$ we need to find the mean first passage times $\tau^*_n$. To do so, we make use of the discrete-time Kolmogorov's backward equation
 \begin{equation}
 \begin{aligned}
 Q_{0,n}(\tau+\Delta \tau)=a_{n+1} \Delta \tau Q_{0,n+1}(\tau) + b_{0,n-1} \Delta \tau Q_{0,n-1}(\tau)  + (1-a_{n+1} \Delta \tau -b_{n-1} \Delta \tau) Q_{0,n}(\tau),
 \end{aligned}
 \label{eq:backward_ME}
 \end{equation}
\noindent where $Q_{0,n}(\tau)$ is the probability of being extinct at time $\tau$, given that initially there were $n$ cells. The backward equation just states that the total probability of becoming extinct from state $n$ at time $\tau + \Delta \tau$ equals the probability of jumping to the state with $n+1$ cells and then going extinct, plus the probability of jumping to state with $n-1$ cells and then going extinct, plus the probability of staying in the same state and going extinct. The probability density of becoming extinct at time $\tau + \Delta \tau$ is $Q_{0,n}(\tau+\Delta \tau)-Q_{0,n}(\tau)$. Hence, the mean first passage time to extinction, starting from a state of $n$ cells, reads
\begin{equation}
\tau^*_n = \sum_{k=0}^{\infty} k \Delta \tau [Q_{0,n}(k \Delta \tau)-Q_{0,n}((k-1) \Delta \tau)]. 
\end{equation}
\noindent Here, we assume $Q_{0,n}(-\Delta \tau)=0$. To simplify the notation, let us denote with $Q_n(\tau)$ the cumulative probability of becoming extinct at time $\tau$ given that the system started in state with $n$ cells.  Substituting $\tau+\Delta \tau \rightarrow \tau$ in Eq.~(\ref{eq:backward_ME}), we have $Q_n(\tau)=a_{n+1} \Delta \tau Q_{n+1}(\tau-\Delta \tau) + b_{n-1} \Delta \tau Q_{n-1} (\tau-\Delta \tau) + (1-\Delta \tau (a_{n+1}+b_{n-1}))Q_n(\tau-\Delta \tau)$. Subtracting both expressions, multiplying by $\tau = k \Delta \tau$ and integrating over time (which in this case amounts to sum over all $k$), leads to
\begin{equation}
    \begin{multlined}
    \sum_{k=0}^{\infty} k \Delta \tau  [Q_n((k+1) \Delta \tau) -  Q_n(k \Delta \tau)]  = 
     \sum_{k=0}^{\infty} a_{n+1} k (\Delta \tau)^2   [Q_{n+1}(\tau) - Q_{n+1}(\tau-\Delta \tau) ]  +   \\ b_{n-1} k (\Delta \tau)^2 [Q_{n-1}(\tau) - Q_{n-1}(\tau-\Delta \tau) ]  +   
   k (\Delta \tau) (1-\Delta \tau (a_{n+1}+b_{n-1})) [Q_{n}(\tau) - Q_{n}(\tau-\Delta \tau)].   
    \end{multlined}    
\end{equation}
We can rewrite the l.h.s.\ as $\sum_{k=0}^{\infty} (k-1) \Delta \tau [Q_n(k \Delta \tau ) - Q_n((k-1) \Delta \tau) ] = \tau_n - \Delta \tau$, where we have used that $Q_n(t)= 0 \; \forall \tau<0$, and the normalization condition of the probability density. The r.h.s.\  of the expression is $a_{n+1} \Delta \tau \tau^*_{n+1} + b_{n-1} \Delta \tau \tau^*_{n-1} + (1- \Delta \tau (a_{n+1} + b_{n-1})) \tau^*_n$. Thus, we obtain
\begin{equation}
    a_{n+1} \tau^*_{n+1} + b_{n-1} \tau^*_{n-1} -(a_{n+1} + b_{n-1}) \tau^*_n + 1 = 0.
\end{equation}
The first boundary condition for this recurrence relation is $\tau^*_0 =0$, i.e., the mean first passage time to extinction starting from extinction is zero. The second condition is $\tau^*_N = \tau^*_{N-1} + 1/b_{N-1}$, i.e., the mean first passage time to extinction from $N$ cells is the corresponding one from $N-1$ cells plus the mean time in which the system hops from $N$ cells to $N-1$ cells. 

To solve for $\tau^*_n$ we define $v_n=\tau^*_n - \tau^*_{n-1}$. Thus, the recurrence relation becomes 
\begin{equation}
\begin{multlined}
a_{n+1}v_{n+1}-b_{n-1} v_n + 1 = 0, \\
    v_N=\frac{1}{b_{N-1}},
\end{multlined}
\end{equation}
which can be solved iteratively (applying the corresponding boundary condition), to yield
\begin{equation}
    v_{n-k}=\sum_{j=1}^{k+1} \frac{1}{b_{N-j}} \sum_{i=j}^k \frac{a_{N-i+1}}{b_{N-i-1}}.
\end{equation}
Finally, solving for $\tau_n^*$ yields
\begin{equation}
    \tau^*_n= \sum_{k=N-n}^{N-1} \sum_{j=1}^{k+1} \frac{1}{b_{N-j}} \prod_{i=j}^k \frac{a_{N-i+1}}{b_{N-i-1}}.
    \label{eq:extinctionTimes}
\end{equation}

Naturally, the accuracy of the expected extinction time $\mathbb{T}$ (and its inverse, $\lambda_1$), is sensitive to the accuracy of the approximate PDF conditioned on non-extinction, $f(n)$.
When calculating $\lambda_1$ from Eq.~(\ref{eq:meanExtinctionTime}) and Eq.~(\ref{eq:extinctionTimes}) (with $f(n)$ estimated numerically), the result is in very good agreement with $\lambda_1$ numerically calculated from the master operator's eigenvalues (compare the orange and black dashed lines in Figure \ref{fig:extinctionTimes}{\bf B}). The approximate $\lambda_1$ calculated using Eq.~(\ref{eq:extinctionTimes}), where $f(n)$ is obtained using one of the approximations for the probability distribution conditioned on non-extinction (Fig.~\ref{fig:extinctionTimes} {\bf B}), is much closer to its real value than $\lambda_1$ calculated via Eq.~\eqref{eq:meanExtinctionTime} (Fig.~\ref{fig:extinctionTimes} {\bf A}).  For high $\phi^*$, all $f(n)$ approximations lead to accurate $\lambda_1$ estimations, with the exception of the half-Gaussian. On the other hand, the half-Gaussian approximation performs well for very low $\phi^*$, when every other approximation of $f(n)$ yields inaccurate $\lambda_1$ estimations. In the intermediate region, the renormalized-system size expansion turns out to be the best choice. 

Extinction times vary significantly between parameter regimes, which we can think of as corresponding to different stem cell niches. Assuming that the system is in quasi-steady state, the average number of stem cells is roughly given by $\langle n \rangle \approx N \phi^*$. Hence, given the average number of stem cells in a niche, we can estimate the effective carrying capacity for different $\phi^*$ values from the quasi-steady state parameter regime. For example, for adult human crypt stem cell niches harbouring $4-6$ cells with division period of $24-30 h$ \cite{barker2009crypt,umar2010intestinal}, the extinction times fall between $13$ days and $50$ years when comparing the results for different $\phi^*$ values spanning all the quasi-steady state region. A predicted extinction time on this scale could point to the need of additional regulatory mechanisms preventing extinction of a tissue's stem cell population within the lifetime of an organism. For slow-cycling stem cells such as haematopoietic stem cells, the lowest estimates yield the presence of $10^4$ cells dividing once every $40$ weeks \cite{catlin2011replication}. More recent estimates point to the number of hematopoietic stem cells in the human body falling within $5\times10^4-2\times10^5$ \cite{lee2018population}. It is unclear how these stem cells are distributed in individual niches, but even for niches hosting $100$ stem cells the extinction times fall within $10^{11}-10^{13}$ years, much higher than the human lifetime. For niches hosting a larger number of stem cells the extinction time is even higher. 

\section{\label{sec:Discussion}Discussion and Conclusion}

We have introduced the birth-death process with volume exclusion, a variation of the birth-death process that incorporates crowding effects due to the finite size of stem cell niches. For cell division to occur there needs to be free space in the niche to accommodate the new born cell --hence the effective proliferation rate is higher (lower) when the niche is less (more) populated. In effect, the expected stem cell number in the vBD model is independent of the initial condition, in contrast to the CBD case, in which the expected number of stem cells is constant in time. Regulation through volume exclusion could also affect clone size distributions and their scaling with average clone size. In contrast to the CBD process, in which the size evolution of different clones is independent, in the vBD all clones are coupled through the empty space species. When one clone grows in size, the probability of other clones growing is decreased, thus affecting the clone-size distribution. Hence, it might be possible to detect evidence of volume exclusion effects from snapshots of clone size distributions, which are commonly measured experimentally \cite{buckingham2011tracing,kretzschmar2012lineage,kester2018single,blanpain2013unravelling}. We will explore this in follow-up research.       

At the stochastic level, the predictions of the vBD master equation differ significantly from those of its deterministic counterpart. While the deterministic rate equations feature a stable steady state, $\phi^*$, to which the system converges logistically, the master equation's solution predicts the vBD model converging to extinction for all parameter sets. However, for $\phi^*$ sufficiently large, a long-lived quasi-steady state appears, and convergence to extinction is very slow. Hence, for high $\phi^*$, a single stochastic trajectory representing a real system, might fluctuate around the quasi-steady state's non-trivial mode for all its lifetime. 

We have shown the vBD model to have three phases with different behaviours that, to the best of our knowledge, have not been analysed before: fast extinction dynamics through a unimodal transient, fast extinction dynamics through a bimodal transient, and slow extinction dynamics with a quasi-steady state. Transient bimodality rarely occurs in chemical reaction networks, but has been reported in recent works \cite{jia2020dynamical,holehouse2020stochastic}. For the quasi-steady state region, we have shown two independent approximate solutions to the master equation, the QSS and WKB approximations. The QSS provides a more accurate approximation, whilst the WKB adopts a simpler mathematical expression. Moreover, the QSS approximation allowed us to prove that the position of the non-trivial mode generally differs from the deterministic prediction. For the bimodal extinction region, we have derived an approximate time-dependent solution to the master equation by making use of a renormalized system-size expansion, which is particularly useful for solving master equations of non-linear birth-death processes, but has not been widely applied. Remarkably, the expression obtained by the renormalized SSE is simpler than the one from the regular SSE. 

The vBD model is mathematically similar to susceptible-infected-susceptible (SIS) models in epidemiology \cite{kryscio1989extinction,jacquez1993stochastic,cao2017stochastic,naasell2011extinction}. Most of the stochastic SIS models studies consider a very large carrying capacity, or let it tend to infinity. Here we have instead focused on solving the master equation for low carrying capacities, motivated by the application to stem cell population dynamics in niches that can have a carrying capacity as low as few tens of cells. The approximate solutions to the master equations we present here, thus, can shed light into the behaviour of SIS-like systems when the effect of having a finite carrying capacity becomes evident. In particular, the solutions of the vBD master equation can be interpreted as the time-dependent PDF of the number of infected individuals, and our approximation of the expected extinction time as the population's expected time to recovery. Another similar model can be found when studying the role of positive feedback in cluster formation of signalling molecules \cite{jilkine2011density}. In this case, the transition from quasi-steady to extinction states are interpreted as a switch from clustered to non-clustered states. A difference between such model and the vBD is that transitions to non-clustered to clustered states are allowed, whilst in the vBD model it is not possible to exit the extinction state.

Our study proposes and describes a minimal model for stem cell dynamics with regulation through competition for space. In this context, the system parameters should be taken as effective parameters encompassing many different features. Let us conclude by discussing aspects of biological realism that could be represented explicitly in future extensions of our model. The vBD model assumes Markovian dynamics, which implies exponentially distributed waiting times between consecutive cell divisions. There is increasing evidence showing that the cell-cycle times are not exponentially distributed \cite{perez2020effects,weber2014quantifying,stumpf2017stem,gavagnin2019invasion,chao2019evidence}. The inclusion of realistic cell-cycle time distributions in the vBD model will require further research. Spatially, our model considers cells distributed on a grid, assuming that the cell shapes remain unaltered. This hypothesis ignores the mechanical plasticity of cells, their mechanical response to pressure in different environments, and the irregular geometry of the stem cell niches \cite{bonakdar2016mechanical,li2005stem}. Moreover, here we have modelled cell populations as well-mixed systems. This well-mixing assumption may be inappropriate when a niche is highly occupied. Since our aim is to study the hallmarks of regulation through volume exclusion, we have also neglected other plausible regulatory mechanisms that might be acting in parallel, such as competition for other resources
\cite{kitadate2019competition,jorg2021stem,kitadate2022regulation} or cell-cell communication pathways between more differentiated (for example, transit amplifying) cells and stem cells \cite{lander2009cell,lo2009feedback}. In open niches these more differentiated cells might also compete with stem cells for space \cite{crane2017adult,kokkaliaris2020adult,kitadate2019competition,kitadate2022regulation}. The role of multiple cell types in competition for space and the interaction between different plausible regulatory mechanisms are interesting avenues for future research.





{ \twocolumn
\bibliography{references}}






\funding{LS was supported by Chancellor's Fellowship from the University of Edinburgh. RG was supported by Chancellor’s Fellow PhD Studentship and Edinburgh Global Scholarship from the University of Edinburgh.}



\end{document}